\documentclass[twocolumn,showpacs,preprintnumbers,amsmath,amssymb]{revtex4}

\usepackage{graphicx}
\usepackage{dcolumn}
\usepackage{bm}
\usepackage{epsf}
\usepackage{graphicx}
\usepackage{amsmath}
\usepackage{subfigure}

\newcommand{\order  }{{\cal O}}

\newcommand{\bra    }{\langle}
\newcommand{\ket    }{\rangle}
\newcommand{\Bra    }{\left\langle}
\newcommand{\Ket    }{\right\rangle}
\newcommand{\atanh  }{{\rm{atanh}}}

\newcommand{\bc     }{\mbox{\boldmath$c$}}

\newcommand{\bsigma }{\mbox{\boldmath$\sigma$}}

\newcommand{\btau   }{\mbox{\boldmath$\tau$}}

\newcommand{\bpsi   }{\mbox{\boldmath$\psi$}}

\newcommand{\tr     }{\textnormal{tr}\:}
\newcommand{\Tr     }{\textnormal{Tr}\:}

\begin{document}

\title{Thermodynamics of spin systems on small-world hypergraphs}
\author{D. Boll\'e, R. Heylen and N.S. Skantzos}
\affiliation{Katholieke Universiteit Leuven, Instituut voor
Theoretische Fysica, Celestijnenlaan 200D, B-3001 Leuven, Belgium}
\email{{desire.bolle,rob.heylen,nikos.skantzos}@fys.kuleuven.be}


\begin{abstract}
We study the thermodynamic properties of spin systems on small-world
hypergraphs, obtained by superimposing sparse Poisson random graphs
with $p$-spin interactions onto a one-dimensional Ising chain with
nearest-neighbor interactions. We use replica-symmetric
transfer-matrix techniques  to derive a set of fixed-point equations
describing the relevant order parameters and free energy,
and solve them employing population dynamics. In the special
case where the number of connections per site is of the order of the
system size we are able to solve the model analytically. In the
more general case where the number of connections is finite we
determine the static and dynamic ferromagnetic-paramagnetic
transitions using population dynamics. The results are
tested against Monte-Carlo simulations.
\end{abstract}
\maketitle

\section{Introduction}

In recent years, a large  amount of work has been devoted to the
study of small-world networks, mainly numerical \cite{pekalski} with
emphasis e.g.\@ on biophysical networks
\cite{girvan}-\cite{barabasi_nature} or social networks
\cite{newman} and, to a lesser extent, analytically
\cite{barratw,reptrans}. For recent reviews see e.g.\@
\cite{albert-barabasi}- \cite{bookbarabasi}.  By now, it has thus
become apparent that small-world architectures can be found in many
different circumstances, ranging from linguistic, epidemic and
social networks to the world-wide-web.

Efficient modeling of real-world applications not only often requires a
diluted random graph to describe the interaction network but,
moreover, these interactions sometimes couple $k$-plets of agents.
For instance, it has been found that the proteomic network of yeast 
forms a hypergraph with the proteins corresponding to
vertices and the protein-complexes corresponding to hyperedges
\cite{ramadan}. Other large metabolic networks like the one of \emph{E.
coli} have also been found to possess a small-world structure and as
most of the reactions of metabolism are multi-molecular they can be
represented by hypergraphs \cite{wagner}. Technical analysis of
sparse hypergraphs is, however, involved even without superimposing
small-world architectures.

A convenient way to describe the statistical physics of this type of
systems is to consider diluted random graphs with $p$-spin
interactions. In this context, a ferromagnetic model having $3$-spin
interactions and finite connectivity has been considered recently in
\cite{ferroglass}. (See also \cite{barratz,ricci}).

A complete description of a small-world system  requires in addition
local interactions. A simple example of the latter is the
nearest-neighbor ferromagnetic Ising interaction. The inclusion of
such local interactions can completely change the functioning and
the dynamics of such systems. It was shown in \cite{reptrans}, e.g.,
that this construction significantly enlarges the region in
parameter space where ferromagnetism occurs. In particular, for any
choice of the value of the average connectivity, however small, the
ferromagnetic-paramagnetic transition occurs at a finite
temperature. Furthermore, a jump in the entropy of metastable
configurations has been found \cite{heylen} exactly at the crossover
between the small-world and the Poisson random graph structure due
to the formation of disconnected clusters within the graph.

In this work we study the thermodynamic properties of such a
small-world hypergraph, obtained by superimposing sparse Poisson
random graphs with $p$-spin interactions onto a one-dimensional
Ising chain with nearest-neighbor interactions. An analytic study of
this model is non-trivial. The relevant disorder-averaged free
energy and order parameters are calculated using replica-symmetric
transfer-matrix techniques. A set of fixed-point equations for the
order parameter functions is derived and solved numerically with the
population dynamics algorithm \cite{mezardparisi}. For $p=2$ we find
back some of the results described in \cite{reptrans},
\cite{1infty,kardar},  for $p \geq 3$ the physics is different. In
the limit where the number of long-range short-cuts are of the order of
the system size for each site we get a mean-field model for which
we are able to solve the order parameters in a completely analytic
way, again using transfer matrices. First-order phase transitions
from the ferromagnetic to the paramagnetic phase are found for all
values of $p$, along with metastable (spinodal)
transitions. Additionally, for $p = 2$ a second-order phase
transition and a coexistence region between the two phases are
found.

Monte-Carlo simulations of the system employing Glauber
dynamics allow us to find the dynamic (metastable)
transition-lines. Very good agreement with the theory is found
when the Ising chain interactions are ferromagnetic. When they are
anti-ferromagnetic we find that the dynamics becomes very slow,
indicating critical slowing down in this region. The rest
of this paper is organized as follows. In Sec.~\ref{secmodel} we
define the small-world model. In Sec.~\ref{secsaddle} we derive the
saddle-point equations for the relevant order parameter function.
Sec.~\ref{sectransfer} discusses the transfer-matrix analysis in the
replica symmetric approximation leading to  a set of fixed-point
equations involving the eigenvectors. Expressions for the free
energy and the physical order parameters are given in
Sec.~\ref{secthermo}. There, we also discuss the continuous
bifurcations from zero magnetization. Sec.~\ref{sec1inf} gives the
analytic solution of the fully connected version of the model. In
Sec.~\ref{secresults} we numerically study the fixed-point equations
and compare the results with simulations. Finally,
Sec.~\ref{secdiscussion} contains the concluding remarks.

\section{The model}\label{secmodel}

\begin{figure}[t]
\vspace{2mm}
\begin{center}
\includegraphics[width=.350\textwidth,height=.350\textwidth]{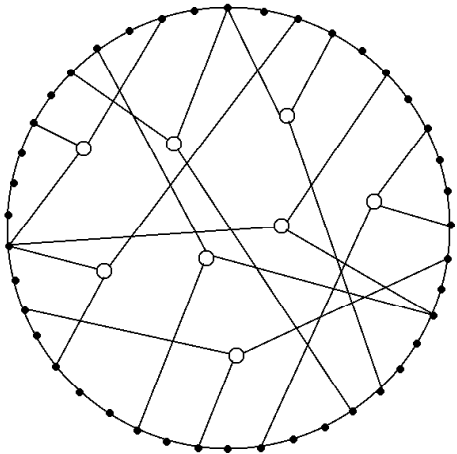}
\caption{Schematic representation of a hypergraph of degree $3$ superimposed onto an Ising chain. Each black dot represents a spin, whereas each circle represents a hyperedge involving  3 spins}
\label{fig:structure}
\end{center}
\end{figure}

Consider a system of $N$ Ising spins $\bsigma = (\sigma_1, \ldots, \sigma_N)$, with $\sigma_i \in \{-1,1\}$, arranged on a one-dimensional chain ($\sigma_{N+1}=\sigma_1)$. Two different couplings are assumed to be present in this system: first, nearest-neighbor interactions of uniform strength $J_0$ and, secondly, sparse long-range $p$-spin interactions of the form $c_{j_1,\ldots,j_{p}}\sigma_{j_1} \ldots \sigma_{j_{p}}$, $\forall j_\ell \in \{1,2, \ldots, N\}$, of uniform strength $J$, which can be described by a hypergraph of degree $p$. An example of such a system for $p=3$ is shown in Fig.~\ref{fig:structure}.

The couplings $c_{j_1, \ldots , j_p}$ are independent identically distributed random variables,   for $j_1 < j_2 < \ldots < j_p$, taken from the following distribution
\begin{eqnarray}
P(c_{j_1, \ldots , j_p}) &=&
    c\ \frac{(p-1)!}{N^{p-1}}\ \delta_{c_{j_1, \ldots , j_p}, 1} \nonumber \\
    &&+ \left(1 - c\ \frac{(p-1)!}{N^{p-1}}\right)\ \delta_{c_{j_1, \ldots , j_p},0}
\label{cdist}
\end{eqnarray}
The value  $c_{j_1, \ldots , j_p} = 1$ indicates that the hyperedge formed by the $p$ spins $(\sigma_{j_1}, \ldots, \sigma_{j_p})$ is present, whereas $c_{j_1, \ldots , j_p} = 0$ means that there is no such hyperedge present.
The quantity $c$ indicates the total number of hyperedges a spin $\sigma$ is, on average, part of
\begin{equation}
\frac{1}{N}\sum_i \sum_{j_{1} < \ldots < j_{p-1}} c_{i,j_1, \ldots, j_{p-1}} = c
\end{equation}
In the small-world context one takes $c$ to be a small number of order ${\cal O}(1)$ while $c/N \rightarrow 0$.

We assume that the couplings are symmetric such that for any permutation $\pi$
in $\mathcal{S}_p$ (the symmetric group of $p$ elements):
\begin{equation}
c_{j_1, \ldots , j_p} = c_{j_{\pi(1)}, \ldots , j_{\pi(p)}}
\end{equation}
Furthermore, we exclude self-interactions and hyperedges of reduced degree by stating that
\begin{equation}
\forall k \neq l \in \{1, \ldots, p\}: j_k = j_l \Rightarrow c_{j_1, \ldots , j_p} = 0
\end{equation}
meaning that the hyperedge does not exist when any two indices are equal.

At thermal equilibrium, such a system can be described by the Hamiltonian
\begin{equation}
H(\bsigma) = -\sum_i \sigma_i h_i(\bsigma) \label{ho}
\end{equation}
where the local field consists of a hypergraph part and a chain part
\begin{eqnarray}
h_i(\bsigma)
  &=&
    \frac{J}{cp}\sum_{j_1 < \ldots < j_{p-1}} c_{i,j_1,\ldots,j_{p-1}}
             \sigma_{j_1} \ldots \sigma_{j_{p-1}} \nonumber \\
   &&+ \frac{J_0}{2}(\sigma_{i-1} + \sigma_{i+1}) \label{hfield}
\end{eqnarray}

We want to study the thermodynamic properties of this network
structure which follow from the free energy per site
\begin{equation}
f=-\lim_{N\to\infty}\frac{1}{\beta N}\log \sum_{\bsigma} e^{-\beta
H(\bsigma)}
\end{equation}
where $\beta =1/T$ corresponds to the inverse bath temperature.

\section{Replicated transfer-matrix analysis}

\subsection{Saddle-point equations}\label{secsaddle}

We start from the free energy per spin written down in the replica approach \cite{MPV}
\begin{eqnarray}
-\beta f(\beta) &=&
   \lim_{N\to\infty}\,\, \lim_{n\to 0}\frac1{Nn}
   \log\bra Z^n\ket_{\bc} \label{eq:fe0} \\
 \bra Z^n\ket_{\bc} &=&
    \sum_{\bsigma^1, \ldots, \bsigma^n}
   \Bra \exp\left(-\beta\sum_{\alpha} H(\bsigma^{\alpha})\right) \Ket_{\bc}
\end{eqnarray}
with $\alpha=1, \ldots, n$ the replica index and the average taken over all possible graphs $\bc$ according to the distribution (\ref{cdist}), so that we obtain
\begin{eqnarray}
\lefteqn{\bra Z^n \ket_{\bc}
 = \sum_{\bsigma^1, \ldots, \bsigma^n}
       \exp \left(\beta J_0 \sum_{i, \alpha}\sigma_i^{\alpha}\sigma_{i+1}^{\alpha}\right)} \nonumber \\
   &\times& \prod_{j_1 < \ldots < j_{p}}
    \Bra \exp \left( \frac{\beta J}{c} \sum_{\alpha} c_{j_1,\ldots,j_{p}}
     \sigma_{j_1}^{\alpha} \ldots \sigma_{j_{p}}^{\alpha} \right)
                    \Ket_{c_{j_1,\ldots,j_{p}}} \\
&=& \sum_{\bsigma^1, \ldots, \bsigma^n}
\exp \left(\beta J_0 \sum_{i, \alpha}\sigma_i^{\alpha}\sigma_{i+1}^{\alpha}\right) \nonumber \\
 &\times& \exp \left( \frac{c\ (p-1)!}{N^{p-1}} \sum_{j_1 < \ldots < j_{p}}
 \left( e^{  \frac{\beta J}{c} \sum_{\alpha} \sigma_{j_1}^{\alpha} \ldots
      \sigma_{j_{p}}^{\alpha} } -1 \right) \right) \label{saddle2}
\end{eqnarray}
where we have used the fact that $N \rightarrow \infty$ to contract the average over $c_{j_1,\ldots,j_{p}}$ into an exponential.

The next step is to insert unities $1=\sum_{\bsigma} \delta_{\bsigma,\bsigma_i}$ and
$1=\sum_{\btau} \delta_{\btau,\bsigma_j}$ where $\bsigma, \btau$ are auxiliary vectors in replica space to arrive at

\begin{eqnarray}
\lefteqn{\bra Z^n \ket_{\bc}
= \sum_{\bsigma^1, \ldots, \bsigma^n}
\exp \left(\beta J_0 \sum_{i, \alpha}\sigma_i^{\alpha}\sigma_{i+1}^{\alpha}\right)} \nonumber \\
&\times& \exp \left( \frac{c}{p N^{p-1}} \sum_{\btau_1 \ldots \btau_p}
   \prod_{k=1}^{p}\sum_{j_k}\delta_{\btau_k, \bsigma_{j_k}} \right. \nonumber \\
&\times& \left. \left( e^{  \frac{\beta J}{c} \sum_{\alpha} \tau_1^{\alpha} \ldots \tau_p^{\alpha} } -1 \right) \right. \Bigg)
 \label{partition1}
\end{eqnarray}

In this way we have effectively introduced an order function
\begin{equation}
F(\btau) = \frac{1}{N}\sum_{i} \delta_{\btau,\bsigma_{i}}
\end{equation}
that can be inserted in (\ref{partition1}) in the usual way
\begin{equation}
1=\int \prod_{\btau}dF(\btau)d\hat{F}(\btau)
\ e^{ i \hat{F}(\btau)
\left(F(\btau)-\frac{1}{N}\sum_{i} \delta_{\btau,\bsigma_i}  \right)}
\end{equation}
to obtain
\begin{eqnarray}
\lefteqn{\bra Z^n \ket_{\bc}
\sim \int \left[\prod_{\btau} dF(\btau)d\hat{F}(\btau)\right]
\exp\left[iN\sum_{\btau}\hat{F}(\btau)F(\btau)\right]} \nonumber \\
&\times&
\exp \left( \frac{cN}{p} \sum_{\btau_1 \ldots \btau_p} \prod_{k=1}^p F(\btau_k)
\left( e^{  \frac{\beta J}{c}\sum_{\alpha} \tau_1^{\alpha} \ldots \tau_p^{\alpha} } -1 \right) \right) \nonumber \\
&\times&\sum_{\bsigma^1, \ldots, \bsigma^n}
 \exp \left(\beta J_0 \sum_{i,\alpha}\sigma_i^{\alpha}\sigma_{i+1}^{\alpha}
  - i\sum_i\hat{F}(\bsigma_i)\right)
\end{eqnarray}
We can then apply the saddle-point method resulting in
\begin{eqnarray}
\lefteqn{\log \left( \bra Z^n \ket_{\bc} \right) =
\underset{F,\hat{F}}{\rm Extr} \left[ i\sum_{\btau} \hat{F}(\btau) F(\btau) \right.}\nonumber \\
&+&\left.  \frac{c}{p}\sum_{\btau_1 \ldots \btau_p} \prod_{k=1}^p F(\btau_k)
\left( e^{  \frac{\beta J}{c}\sum_{\alpha} \tau_1^{\alpha} \ldots \tau_p^{\alpha} } -1 \right)
\right. \label{logZ} \nonumber \\
&+&\left.
 \frac{1}{N} \log \left( \sum_{\bsigma^1, \ldots, \bsigma^n}
e^{ \beta J_0 \sum_{i\alpha}\sigma_i^{\alpha}\sigma_{i+1}^{\alpha}
 -i\sum_i\hat{F}(\bsigma_i)} \right) \right]
\end{eqnarray}

Derivation with respect to $F(\bpsi)$ and $\hat{F}(\bpsi)$ results in the following self-consistent equation for the density $F(\psi)$:
\begin{widetext}
\begin{equation}
F(\bpsi) = \frac
   {\sum_{\bsigma_1 \ldots \bsigma_N}
        \left[\frac{1}{N}\sum_i \delta_{\bsigma_i, \bpsi}\right]
      \exp \left( \beta J_0 \sum_{j\alpha} \sigma_j^{\alpha} \sigma_{j+1}^{\alpha}
           + c \sum_j \sum_{\btau_1 \ldots \btau_{p-1}} \prod_{k=1}^{p-1} F(\btau_k)
   \left( e^{  \frac{\beta J}{c}\sum_{\alpha} \tau_1^{\alpha} \ldots \tau_{p-1}^{\alpha}
                       \sigma_j^{\alpha} } -1 \right)\right)}
   {\sum_{\bsigma^1 \ldots \bsigma^n}
    \exp \left( \beta J_0 \sum_{j\alpha} \sigma_j^{\alpha} \sigma_{j+1}^{\alpha}
      + c \sum_j  \sum_{\btau_1 \ldots \btau_{p-1}} \prod_{k=1}^{p-1} F(\btau_k)
   \left( e^{  \frac{\beta J}{c}\sum_{\alpha} \tau_1^{\alpha} \ldots \tau_{p-1}^{\alpha}
 \sigma_j^{\alpha} } -1 \right)\right)} \label{F}
\end{equation}
\end{widetext}
In the absence of short-range bonds, i.e.\@ $J_0=0$, this expression
factorizes over sites and can be reduced considerably. In our case,
however, this is not possible and in order to perform the spin
summations we are now required to construct transfer matrices.

\subsection{Transfer-matrix analysis}\label{sectransfer}

Defining the following $2^n \times 2^n$ matrix
\begin{eqnarray}
\lefteqn{T_{\bsigma,\btau}[F]
= \exp \left( \beta J_0 \sum_{\alpha} \sigma^{\alpha} \tau^{\alpha} \right)} \nonumber \\
&\times& \exp \left( c \sum_{\btau_1 \ldots \btau_{p-1}} \prod_{k=1}^{p-1} F(\btau_k)
\left( e^{  \frac{\beta J}{c}\sum_{\alpha} \tau_1^{\alpha} \ldots \tau_{p-1}^{\alpha} \sigma^{\alpha} } -1 \right)\right) \nonumber \\ \label{T}
\end{eqnarray}
Eq.~(\ref{F}) reads
\begin{eqnarray}
\lefteqn{F(\bpsi) = } \nonumber \\
&&\frac
{\sum_j\sum_{\bsigma_1 \bsigma_j} \left( T^{j-1}[F] \right)_{\bsigma_1 \bsigma_j}
 \delta_{\bsigma_j, \bpsi} \left( T^{N-j+1}[F] \right)_{\bsigma_j \bsigma_1}}
{N\ \tr (T^N[F])} \nonumber \\
\end{eqnarray}
We insert unity  $1 = \sum_{\btau} \delta_{\bsigma_j, \btau}$ and introduce the matrix $Q_{\bsigma_j \btau}(\bpsi) = \delta_{\bsigma_j, \bpsi} \delta_{\bsigma_j, \btau}$ to
obtain  after some algebra
\begin{equation}
F(\bpsi) = \frac{\tr (T^{N}[F] Q(\bpsi))}{\tr (T^N[F])} \label{F2}
\end{equation}

To proceed with the evaluation of the traces in (\ref{F2}) we remark that in the thermodynamic limit $N\rightarrow\infty$ only $\lambda_0$, the largest eigenvalue of $T[F]$ will contribute. Defining
\begin{eqnarray}
\sum_{\btau} T_{\bsigma,\btau}[F] u(\btau) &=& \lambda_0 u(\bsigma) \label{u}\\
\sum_{\bsigma} v(\bsigma) T_{\bsigma,\btau}[F]  &=& \lambda_0 v(\btau) \label{v}
\end{eqnarray}
we have that
\begin{equation}
T^N_{\bsigma,\btau}[F] \approx \lambda_0^N u(\bsigma) v(\btau)
\end{equation}
and consequently
\begin{equation}
F(\bpsi) = \frac{u(\bpsi) v(\bpsi)}{\sum_{\bsigma} u(\bsigma)
v(\bsigma)} \label{Fuv}
\end{equation}
So, in order to find a solution for $F(\bpsi)$ we need to solve the equations (\ref{u}), (\ref{v}) and (\ref{Fuv}).

At this point we invoke replica symmetry (RS) by assuming that
$
\forall \pi \in \mathcal{S}_n : F(\bpsi) = F(\pi(\bpsi)).
$
One way to fulfil this is to write $F(\bpsi)$ as follows \cite{mezardparisi}
\begin{equation}
F(\bpsi) = \int dh\: W(h) \prod_{\alpha=1}^n \frac{e^{\beta h \psi^\alpha}}{2 \cosh (\beta h)} \label{FRS}
\end{equation}
The density $W(h)$ is normalised. We also assume the left and right eigenvectors to be replica symmetric
\begin{eqnarray}
u(\bpsi) &=& \int dx\: \phi(x) \prod_{\alpha=1}^n e^{\beta x \psi^\alpha} \label{uRS} \\
v(\bpsi) &=& \int dy\: \chi(y) \prod_{\alpha=1}^n e^{\beta y \psi^\alpha} \label{vRS}
\end{eqnarray}

This allows us to write down self-consistent equations for $\phi(x)$
and $\chi(x)$. We insert Eqs.~(\ref{FRS}) and (\ref{uRS}) into the
l.h.s of Eq.~(\ref{u}) and obtain after some algebra (see the
Appendix for more details) the following closed equation for $n
\rightarrow 0$
\begin{eqnarray}
\lefteqn{\lambda_0 \phi(x')
=
\sum_{\mu=0}^{\infty} \left\{\frac{e^{-c}c^{\mu}}{\mu!}
\left[ \prod_{\nu=1}^{\mu}\prod_{k=1}^{p-1}\! \int\! dh_k^{\nu}\: W(h_k^{\nu}) \right]\! \int\! dx\: \phi(x) \right.}
\nonumber \\
&&\times \left.
\delta\left[ x'
- \frac{1}{\beta} \left[
\sum_{\nu=1}^\mu \atanh \left(\tanh (\frac{\beta J}{c}) \prod_{k=1}^{p-1} \tanh (\beta h_k^\nu)\right) \right.\right.\right. \nonumber \\
&& \left.  \left.+ \atanh \left(  \tanh(\beta x) \tanh(\beta J_0) \Bigg{)} \right] \right]
\right\} \label{phi}
\end{eqnarray}

In a similar way we derive a self-consistent equation for $\chi(x)$ by inserting (\ref{FRS}) and (\ref{vRS}) into the l.h.s. of Eq.~(\ref{v})
\begin{widetext}
\begin{eqnarray}
\lambda_0 \chi(x')
&=&
\sum_{\mu=0}^{\infty} \left(\frac{e^{-c}c^{\mu}}{\mu!}
\left[ \prod_{\nu=1}^{\mu}\prod_{k=1}^{p-1} \int dh_k^{\nu}\: W(h_k^{\nu}) \right] \right.
\nonumber \\
&& \left.
\int dx\: \chi(x) \delta\left[ x'
- \frac{1}{\beta} \atanh \left[\tanh(\beta J_0) \tanh \left( \beta x + \sum_{\nu=1}^\mu \atanh ( \tanh(\frac{\beta J}{c}) \prod_k \tanh(\beta h_k^\nu )) \right) \right] \right]
\right\} \label{chi}
\end{eqnarray}
\end{widetext}
At this point  we choose the $\phi(x)$ and $\chi(x)$ to be
normalised. We remark that in the limit $c \rightarrow 0$, $J
\rightarrow 0$ or $p \rightarrow \infty$ equations (\ref{phi}) and
(\ref{chi}) reduce correctly to those of a one-dimensional Ising
chain.

Next, to find  the self-consistent equation for $W(h)$ we start from
Eq.~(\ref{Fuv}) and fill in the RS-assumptions (\ref{uRS}) and
(\ref{vRS}). Requiring that the resulting expression takes in
the limit $n\rightarrow 0$ the form of Eq.~(\ref{FRS}) we obtain
\begin{equation}
W(h) = \int dx\: dy\: \phi(x) \chi(y) \delta(h -x -y) \label{Wsc}
\end{equation}

Finally, in order to calculate  the free energy per spin we need to
determine the largest eigenvalue $\lambda_0$ in the limit
$n\rightarrow 0$. We start from Eq.~(\ref{u}), insert
Eqs.~(\ref{FRS}) and (\ref{uRS}) to obtain
\begin{eqnarray}
\lefteqn{\lambda_0 =
1 + n \sum_{\mu=0}^{\infty} \frac{e^{-c}c^{\mu}}{\mu!}
\prod_{\nu=1}^{\mu}\prod_{k=1}^{p-1} \int dh_k^{\nu}\: W(h_k^{\nu}) \int dx\: \phi(x)
}\nonumber \\
&& \times \left( \frac{1}{2} \sum_s \left(\log \left( G^{R}_{s}(x,\{h_k^{\nu}\}) \right)
  \right)
- \log(2 \cosh ((\beta h_k^{\nu})))\right) \nonumber \\
&&+ \order (n^2) \label{lambda0}
\end{eqnarray}
with
\begin{eqnarray}
\lefteqn{G^{R}_{s}(x,\{h_k^{\nu}\}) =
\left( \sum_{\gamma=\pm 1}e^{\beta \gamma (x + J_0  s)}\right) } \nonumber \\
&\times& \prod_{\nu=1}^{\mu} \sum_{\gamma_1 \ldots \gamma_{p-1}}
e^{\beta \sum_{k=1}^{p-1} h_k^{\nu} \gamma_k + \frac{\beta J}{c} \gamma_1 \ldots
\gamma_{p-1} s} \label{GRS}
\end{eqnarray}
So $\lambda_0 = 1$ in the limit $n\rightarrow 0$.

\subsection{Thermodynamics}\label{secthermo}

We can now evaluate the free energy per spin. Starting from (\ref{eq:fe0}) and (\ref{logZ}) we arrive at
\begin{widetext}
\begin{eqnarray}
-\beta f(\beta) &=&
\frac{c(1-p)}{p} \left[ \prod_{k=1}^p \int dh_k\: W(h_k) \right]
\left[\log \left( \sum_{\tau_1 \ldots \tau_p}
e^{\beta \sum_k h_k \tau_k + \frac{\beta J}{c} \tau_1 \ldots \tau_p} \right)
   - \log \left( \sum_{\tau_1 \ldots \tau_p}e^{\beta \sum_k h_k \tau_k } \right) \right]
   \nonumber \\
&& + \sum_{\mu=0}^{\infty} \frac{e^{-c}c^{\mu}}{\mu!}
    \prod_{\nu=1}^{\mu}\prod_{k=1}^{p-1} \int dh_k^{\nu}\: W(h_k^{\nu}) \int dx\: \phi(x)
    \left( \frac{1}{2} \sum_s
      \left(\log \left( G^{R}_{s}(x,\{h_k^{\nu}\}) \right)  \right)
     - \log(2 \cosh ((\beta h_k^{\nu})))\right) \label{freeE}
\end{eqnarray}
\end{widetext}
which is the final result. As the
elements of the adjacency matrix $c_{i_1,\ldots,i_p}$ have been
taken i.i.d.\@ one would expect the degrees at each site to be
Poisson distributed. Indeed, we see that this has come out naturally
from the analysis and we can associate the average over the Poisson
probabilities in (\ref{phi},\ref{chi},\ref{freeE}) as an average
over degrees.

The order parameters of the system under study are the average
magnetization which reads, recalling Eq.~(\ref{FRS})
\begin{eqnarray}
m_\alpha &=& \Bra \frac{1}{N} \sum_i \sigma_i^\alpha \Ket_{\bc} \\
&\overset{\text{RS}}{=}& \int dh\: W(h) \tanh(\beta h) \label{mag}
\end{eqnarray}
and the Edwards-Anderson parameter function
\begin{eqnarray}
q_{\alpha\beta} &=&
  \Bra \left(\frac{1}{N} \sum_i \sigma_i^\alpha \sigma_i^\beta\right) \Ket_{\bc} \,\,\,,
   \qquad \alpha \neq \beta \\
&\overset{\text{RS}}{=}& \int dh\: W(h) \tanh^2(\beta h) \label{ea}
\end{eqnarray}
We remark that due to the RS  assumption $m_\alpha = m$ for $\forall
\alpha$, $q_{\alpha\beta}=q$ for $\forall \alpha \neq \beta$, and
$q_{\alpha \alpha}=1$.

Next, in order to obtain the  phase diagram we have to study the
solutions of the self-consistent equations (\ref{phi}), (\ref{chi})
and (\ref{Wsc}). It is easily seen that the field-distributions
$\phi(x) = \chi(x) = W(x) = \delta(x)$ are a solution of these
equations, corresponding to the paramagnetic phase according to
Eq.~(\ref{mag}). For this solution the free energy per spin
(\ref{freeE}) reduces to
\begin{eqnarray}
-\beta f_\text{para}(\beta) &=& \frac{c}{p} \log \cosh \left( \frac{\beta J}{c} \right) \nonumber \\
&&+ \log \cosh \left( \beta J_0 \right) + \log(2) \label{freeE_para}
\end{eqnarray}
For high temperatures, this is the only solution present.
Following a standard procedure in finite-connectivity theory (see,
e.g. \cite{reptrans}) we discuss continuous bifurcations away from
this solution in order to find second-order phase transitions. Since
we are dealing with field-distributions this means that the fields
will be narrowly distributed around zero so that we can expand the
equations (\ref{mag}) and (\ref{ea}), using equation (\ref{Wsc}). In
order to quantify the difference with the deltapeak-solution we
assume $\int dh \: h^k \phi(h) = \order(\epsilon^k)$ and $\int dh \:
h^k \chi(h) = \order(\epsilon^k)$ with $|\epsilon| \ll 1$ such that
\begin{eqnarray}
m &=& \beta \int dx\: \phi(x)x + \beta \int dy\: \chi(y)y + \order(\epsilon^3) \\
q &=& \beta^2 \int dx\: dy \: \phi(x) \chi(y) (x+y)^2  + \order(\epsilon^3)
\end{eqnarray}
In order to find the transition towards non-zero magnetization we
look for bifurcations where the first moments become of order
$\epsilon$. Introducing $\overline{x} = \int dx\: x\: \phi(x)$ and
$\overline{y} = \int dy\: y\: \chi(y)$, and recalling
Eqs.~(\ref{phi}) and (\ref{chi}) we find the following set of
self-consistent equations
\begin{eqnarray}
\overline{x} &=& \overline{x} \tanh (\beta J_0) + c \tanh(\frac{\beta J}{c})  (\overline{x} + \overline{y})^{p-1} \beta^{p-2} \qquad \\
\overline{y} &=& \overline{y} \tanh (\beta J_0) \nonumber \\
&&+ c \tanh (\beta J_0)  \tanh(\frac{\beta J}{c})  (\overline{x} + \overline{y})^{p-1} \beta^{p-2}
\end{eqnarray}
For $p>2$ the second terms in the r.h.s. of these equations are of a higher order in $\epsilon$ and, hence, no second-order  bifurcations to a ferromagnetic phase are found. For $p=2$ these equations do lead to a second-order bifurcation at
\begin{equation}
1 =  c \tanh(\frac{\beta J}{c}) \exp(2\beta J_0) \label{bifm}
\end{equation}
in agreement with \cite{reptrans}.

Analogously, we can look for bifurcations to a spin-glass transition by expanding the second-order moments, assuming that $\overline{x} = \overline{y} = 0$. Introducing  $\overline{x^2} = \int dx\: x^2\: \phi(x)$ and $\overline{y^2} = \int dy\: y^2\: \chi(y)$ we get
\begin{eqnarray}
\overline{x^2} &=& \overline{x^2} \tanh^2 (\beta J_0) \nonumber \\
&& + c  \tanh^2(\frac{\beta J}{c})  \overline{(x + y)^{2(p-1)}} \beta^{p-2} \\
\overline{y^2} &=& \overline{y^2} \tanh^2 (\beta J_0) \nonumber \\
&&+ c\tanh^2 (\beta J_0) \tanh^2(\frac{\beta J}{c}) \overline{(x + y)^{2(p-1)}} \beta^{p-2} \nonumber \\
\end{eqnarray}
Again, from this we can deduce that there is no second-order spin-glass transition for $p>2$ but for $p=2$ a continuous bifurcation to $q > 0, m=0$ occurs at
\begin{equation}
1 =  c \tanh^2(\frac{\beta J}{c})  \cosh(2\beta J_0) \label{bifq}
\end{equation}

The $p=2$ results are in agreement with those of \cite{reptrans},
where it is also argued that for $J, J_0 \geq 0$ the second-order
paramagnetic to spin-glass instability cannot occur as it will
always be preceded by the ferromagnetic one when lowering the
temperature. From simulation experiments we find some evidence for
glassy behavior at lower temperatures for all $p$, especially for
$\beta J_0 \leq 0$, as will be shortly discussed in
Section~\ref{secresults}. There we also study first-order
transitions by employing the transfer-matrix analysis developed in
Sec.~\ref{secsaddle},\ref{sectransfer} together with population
dynamics.

\section{The $1+\infty$ dimensional model}\label{sec1inf}

A limiting case of the previous results is a one-dimensional Ising chain superimposed on a fully connected hypergraph. This model gets a mean-field character and can be solved analytically by using the transfer-matrix approach. In the sequel we denote this model by the $1+\infty$ dimensional model.

In order to have a multispin interaction between every multiplet of $p$ spins we must take  $c = N^{p-1}/(p-1)!$ (See e.g. equation (\ref{cdist})). Recalling the Hamiltonian (\ref{ho}) and the local field (\ref{hfield}) this leads to the following free energy per spin
\begin{widetext}
\begin{eqnarray}
-\beta f &=& \lim_{N \rightarrow \infty} \frac{1}{N} \log
\left(\underset{\bsigma}{\Tr} \exp \left( \frac{\beta J}{p N^{p-1}}\sum_{j_1}\ldots \sum_{j_{p}} \sigma_{j_1} \ldots \sigma_{j_{p}}
    + \beta J_0 \sum_i \sigma_{i}\sigma_{i+1} \right) \right)  \\
&=& \lim_{N \rightarrow \infty} \frac{1}{N} \log
\left( \underset{\bsigma}{\Tr} \int dm d\hat{m} \exp
  \left( i \hat{m} (m - \frac{1}{N} \sum_{i} \sigma_{i}) \right)
    \exp \left(  \frac{\beta J N}{p}  m^p +
           \beta J_0 \sum_i \sigma_{i}\sigma_{i+1} \right) \right) \\
&=& \lim_{N \rightarrow \infty} \frac{1}{N} \log
\left( \int dm d\hat{m} \exp \left( i N \hat{m} m + \frac{\beta J N}{p}  m^{p} \right)
    \Tr \left[ T^N \right] \right) \label{1free1} .
\end{eqnarray}
\end{widetext}
Here $T$ is the transfer-matrix
\begin{equation}
T = \left(
  \begin{array}{cc}
      e^{-i\hat{m} + \beta J_0} & e^{-i\hat{m} - \beta J_0} \\
       e^{i\hat{m} - \beta J_0} & e^{i\hat{m} + \beta J_0}
   \end{array}
   \right)
\end{equation}
The eigenvalues $\lambda$ of this matrix are
\begin{eqnarray}
\lambda_{\pm}(\hat{m})
&=& e^{\beta J_0} \cosh(i\hat{m}) \nonumber \\
       &&\pm \left( e^{2\beta J_0} \cosh^2(i\hat{m})
                    - 2 \sinh(2 \beta J_0) \right)^{\frac{1}{2}}
\end{eqnarray}
Using the fact that for $\to\infty$ $\Tr \left[ T^N \right] =
\lambda_{+}^N + \lambda_{-}^N \approx \lambda_{+}^N$ we write
equation (\ref{1free1}) as
\begin{eqnarray}
-\beta f &=&
\lim_{N \rightarrow \infty} \frac{1}{N} \log
      \left( \int dm d\hat{m} \right. \nonumber \\
             &&\times \left. \exp \left( i N \hat{m} m+ \frac{\beta J N}{p}  m^{p}
                       + N \log(i \lambda_{+} \hat{m})\right) \right) \nonumber \\
\end{eqnarray}
In the limit $N \rightarrow \infty$, the fixed-point equation
minimizing the free energy per spin is given by
\begin{equation}
m = G(m), \hspace{2mm} G(m)\equiv\frac{\sinh (\beta J m^{p-1})}
    {\sqrt{\sinh^2(\beta J m^{p-1}) + e^{-4 \beta J_0}}} \label{1m}
\end{equation}
We remark that this equation reduces to the one presented in \cite{1infty} for $p = 2$.

To find the phase diagram of this system we perform a bifurcation
analysis. It is easy to see that $m = 0$, which describes the
paramagnetic phase, always satisfies Eq.~(\ref{1m}) for any
temperature. To find phase transitions away from the paramagnetic
solution, we must find the critical parameter values for which the
pair of equations $m=G(m)$ and $1=G'(m)$ has new $m\neq 0$
solutions. These solutions can be created from $m=0$ (continuous
bifurcation) or away from $m=0$ (discontinuous ones).

With these considerations we immediately observe that there are no
solutions for $p>2$, but for $p=2$ we find the solution
\begin{eqnarray}
\beta J  = e^{-2 \beta J_0} \label{dis2}
\end{eqnarray}
To look for first-order transitions we have to find solutions to
$m=G(m)$ and $1=G'(m)$ at $m\neq 0$. In this model this can be done
analytically by introducing the auxiliary variable
\begin{equation}
x = \beta J m^{p-1}, \qquad x \in [-\infty, \infty]/\{0\} \label{x}
\end{equation}
leading to the following two equations describing the transition
line, parametrized by $x$:
\begin{eqnarray}
\beta J_0(x) &=& -\frac{1}{4} \log \left( \frac{\tanh(x) \sinh^2(x)}{x(p-1) - \tanh(x)} \right) \label{betaJ0} \\
\beta J(x) &=& {x} \left( \frac{x(p-1)}{x(p-1) - \tanh(x)} \right)^{\frac{1}{2}(p-1)} \label{betaJ}
\end{eqnarray}
We remark that in the limit $\beta J \to \infty$, the slope of $\beta J/ \beta J_0$ approaches $-2$ and, hence, is $p$-independent.

\begin{figure}[ht]
\centering
\includegraphics[width=.45\textwidth]{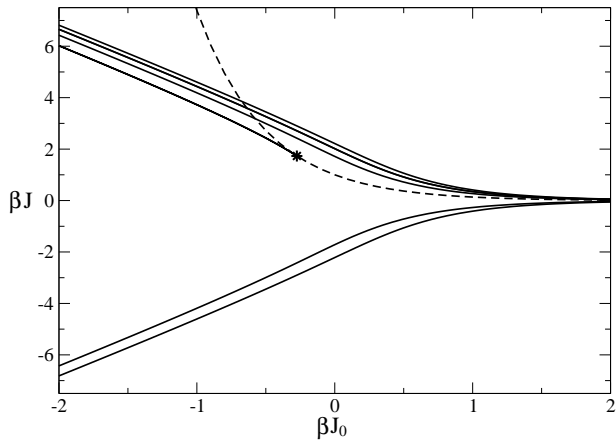}
\caption{Phase diagram for the $ 1+ \infty $ dimensional model in the $\beta J - \beta J_0$ plane for different $p$. The solid lines indicate first-order transitions: for $\beta J > 0$, $p=2,3,4,5$ from bottom to top, and for $\beta J < 0$, $p=3,5$ from top to bottom. The dashed line is the second-order transition line for $p=2$. The star indicates the tri-critical point for $p=2$.}
\label{phase_1inf}
\end{figure}

For any non-zero $x$ we  can now find a pair $\beta J$ and $\beta
J_0$ at which a first order phase transition occurs. The value of
the magnetization $m$ at that point is given by equation (\ref{x}).
These phase transition lines are plotted in Fig.~\ref{phase_1inf}
for several values of the degree $p$.

\vspace*{1cm}
We see that no transitions  occur for even $p$ and $\beta J < 0$.
Furthermore, we remark the $\beta J$ symmetry for odd $p$, due to
the symmetry properties of the Hamiltonian (\ref{ho}) and
(\ref{hfield}) under the change of sign $\bsigma \rightarrow
-\bsigma$. The ferromagnetic phase is situated to the right of the
transition lines, so for odd $p$ the paramagnetic phase is in
between the symmetrical solid lines, for $p$ even it lies below the
corresponding solid line. For increasing $p$, the ferromagnetic
region decreases. For the special case of $p=2$ the paramagnetic and
ferromagnetic phase coexist between the first and second-order
transition lines with as tri-critical point $\beta J = \sqrt{3}
\approx 1.732, \,\, \beta J_0 = -\log(3)/4 \approx -0.275$. The
latter results are in agreement with the results of \cite{1infty}
and with those of \cite{kardar} in the case of one dimension. This
analysis serves as a limiting case of our small-world model
for increasingly larger values of the mean connectivity per site
$c$.

\section{Results for finite $c$.}\label{secresults}

The main equations describing the thermodynamics of the small-world
hypergraph for finite $c$ are Eqs.~(\ref{phi}), (\ref{chi}) and
(\ref{Wsc}). To solve these equations we use the population dynamics
algorithm to generate field distributions together with Monte Carlo
integration over the generated populations in order to obtain the
physical parameters. The important parameters of this algorithm are
the size of the populations and the number of iterations. The size
of the populations has to be big enough to get clearly outlined
distributions keeping in mind, however, that the computational time
required for the algorithm to converge is linear in this parameter.
In most cases we find that populations of 10000 fields give accurate
results. The number of iterations per  spin depends strongly on the
physical parameters. It turns out that most of the time about 1000
iterations results in a reasonable accuracy. To calculate
the ferromagnetic free energy it proves useful to average over
several (e.g. 100) runs with different initial conditions.

From  the $1 + \infty$ dimensional model solved analytically in
Section~\ref{sec1inf} we learned already that the physics for $p=2$
versus the one for $p \geq 3$ might be very different. As a
benchmark test for our derivations we have reproduced some of the
results for $p=2$ found in \cite{reptrans, kardar}. We do not repeat
them here but concentrate on $p \geq 3$ in the sequel.

Starting from the paramagnetic phase and lowering the
temperature, ferromagnetic solutions will start to appear,
indicating a dynamical transition which appears to be first order.
However, to check which of these solutions is thermodynamically
stable, we need to calculate the free energy of both the $m=0$ and
$m\neq 0$ solutions. The former is given by (\ref{freeE_para}), and
the latter can be calculated numerically from (\ref{freeE}). This
leads to two special temperatures: the temperature where the first
$m\neq 0$ solutions start to appear, also known as the spinodal
point, dynamical transition or metastable transition, and the
temperature where the ferromagnetic free energy becomes lower than
the paramagnetic free energy, indicating the thermodynamic phase
transition.

\vspace*{1cm}
\begin{figure}[ht]
\begin{center}
\leavevmode
\includegraphics[width=.45\textwidth]{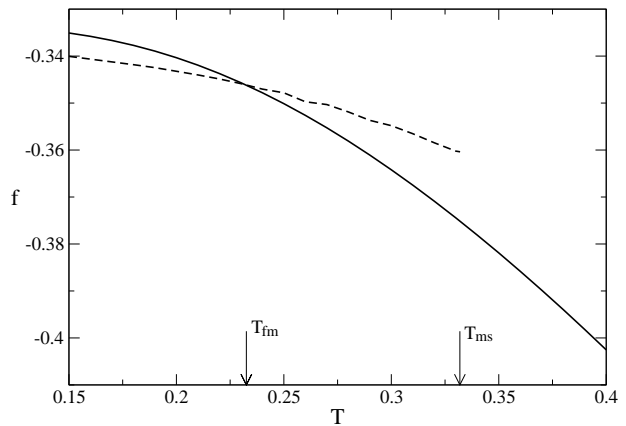}
\caption{The free energy per spin as a function of
temperature for $J=1$, $J_0=0$, $c=3$ and $p=3$. The dashed line
indicates the free energy of the $m\neq 0$ state, while the solid
line indicates the paramagnetic free energy. The thermodynamic phase
transition occurs at the crossing of the two lines, whereas the
spinodal point is the highest temperature for which an $m\neq 0$
solution is possible. } \label{freeE_plot}
\end{center}
\end{figure}

\begin{figure}[ht]
\begin{center}
\leavevmode
\includegraphics[width=.45\textwidth]{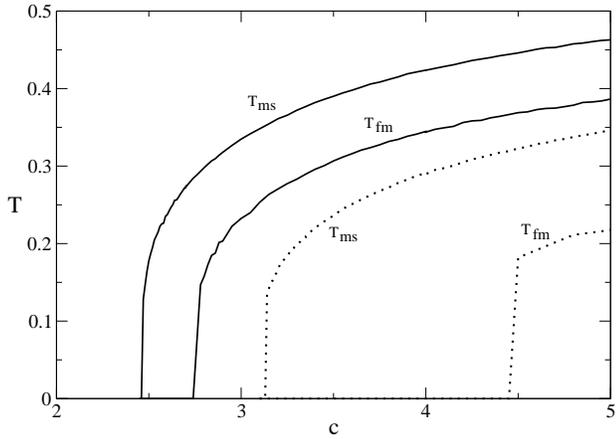}
\caption{Critical temperatures as a function of the average connectivity $c$ for the ferromagnetic-paramagnetic phase transition ($T_{fm}$) and the spinodal line ($T_{ms}$). Parameters are: $p=3$, $J_0=0$ (solid); $p=4$, $J_0=0$ (dotted); For all lines $J=1$.}
\label{tc}
\end{center}
\end{figure}

\begin{figure}[ht]
\begin{center}
\leavevmode
\includegraphics[width=.45\textwidth]{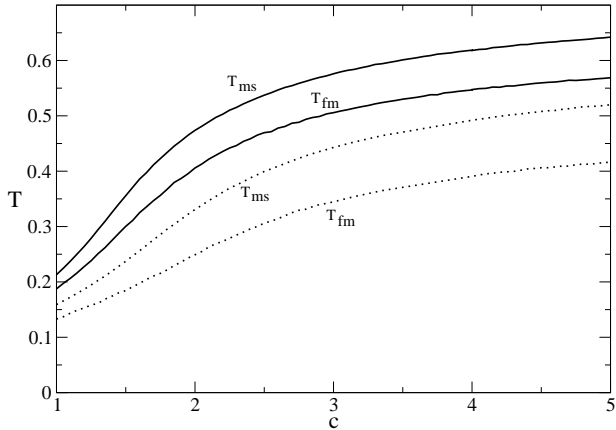}
\caption{Critical temperatures as a function of the average connectivity $c$ for the ferromagnetic-paramagnetic phase transition ($T_{fm}$) and the spinodal line ($T_{ms}$). Parameters are: $p=3$, $J_0=0.1$ (solid); $p=4$, $J_0=0.1$ (dotted); For all lines $J=1$.}
\label{tc2}
\end{center}
\end{figure}

First, we calculate for general $p$ the critical temperatures at
which these transitions occur. They are
plotted for the the cases $p=3,4$ for the model without chain contribution ($J_0 = 0$) in Fig.~\ref{tc} , and with chain contribution ($J_0 = 0.1$) in Fig.~\ref{tc2}. Above the
thermodynamic transition lines ($T_{fm}$) we find the paramagnetic
phase, below the ferromagnetic phase. Metastable ferromagnetic
states can be found up until the corresponding spinodal lines
($T_{ms}$). For all $J_0$ we see that the critical temperature
decreases with increasing $p$. For $J_0 = 0$ a higher $c$ is
required to have ferromagnetic behavior. For numerical reasons we do
not consider very small $T$ or $c$. The results for $p=3$ and
$J_0=0$ are in agreement with the $T_{fm}$ and $T_{ms}$ transition
lines given in \cite{ferroglass} (note that our Hamiltonian is rescaled with a factor c).

\vspace*{1cm}

\vspace*{1cm}

\vspace*{1.3cm}
\begin{figure}[ht]
\begin{center}
\leavevmode
\includegraphics[width=.45\textwidth]{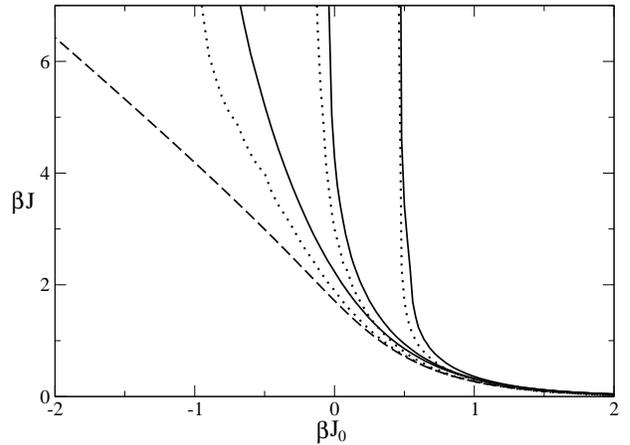}
\caption{Phase transition lines in the $\beta J - \beta J_0$ plane for $p=3$ and different  connectivities $c$: solid lines from right to left $c=1, 3, 10$.  The dotted lines are the corresponding spinodal lines. The dashed line is for $c =  \infty$.}
\label{phase_diffc}
\end{center}
\end{figure}

Furthermore, we search for the transition lines  in the $\beta J -
\beta J_0$ plane. These transitions between $m\neq 0$  and $m=0$
 are plotted in Fig.~\ref{phase_diffc} for $p=3$, $\beta J >
0$ and $c= 1, 3, 10$ (solid lines), together with the
spinodal lines (dotted lines) and the theoretical result for $c
\rightarrow \infty$ (dashed line), which is solved
analytically in Section~\ref{sec1inf}. All transitions shown here
are first-order. The ferromagnetic phase is situated to the right of
the transition lines and increases substantially with growing $c$.
For bigger values of $c$, the transitions approach the analytically
derived $c=\infty$ result. Because $p$ is odd, the transition lines
for $\beta J < 0$ are found by reflection symmetry with respect to
the $\beta J_0$ axis. Just as in the $p=2$ case, the small-world
hypergraph has its ferromagnetic transition at finite temperatures
for all non-zero values of $c$. Analogously to the $1+\infty$
dimensional case (see fig. \ref{phase_1inf}), the ferromagnetic
region decreases with increasing $p$, and disappears for $p
\rightarrow \infty$.

\begin{figure}[ht]
\begin{center}
\leavevmode
\includegraphics[width=.45\textwidth]{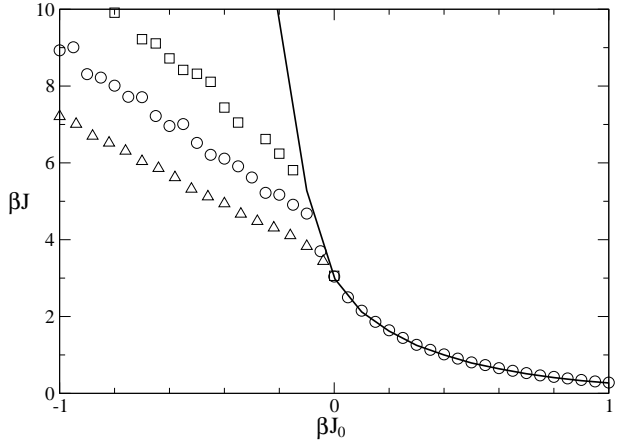}
\caption{Phase transition lines in the $\beta J - \beta J_0$ plane
for $p=3, c=3$. The solid line represents the spinodal line found by
the population dynamics result. The triangles, circles and squares
indicate the results  of simulations with $10^4$ spins and,
respectively, $2000, 5 \times 10^4, 5 \times 10^6$ iterations per
spin.} \label{simspin}
\end{center}
\end{figure}

Simulations have been performed for this small-world model with
heat-bath dynamics and sequential updating. Our results in this perspective are
somewhat limited by the nature of the phase transition: We are
dealing with a very sparse system undergoing a first-order phase
transition. Metastabilities will be present (as already indicated by
the presence of the spinodal lines) and cause slow dynamics near the
thermodynamic transition line. This in turn causes the system to
show strong hysteresis effects. With these simulations we can look
for the spinodal line by initializing the system in a fully
magnetized state, and looking for the temperature where order
disappears at long times. As a typical example the results for $p=3,
c=3$ are plotted in Fig.~\ref{simspin} for $10^4$ spins and
different numbers of iterations. Very good agreement with the
spinodal line obtained with the population dynamics solution is
found in the positive $\beta J_0$ region. When $J_0$ is negative
however we do not find satisfactory results. This can be explained
by the opposing forces at work in the system (ferromagnetic graph
and antiferromagnetic chain), which will only slow the dynamics
further down. We were unable to pinpoint the thermodynamic phase
transitions with simulations due to the effects mentioned above. The
simulations show some further evidence for glassy dynamics (very
large spin-spin autocorrelation times) for lower temperatures, but a
detailed discussion of such non-trivial glassy behavior is beyond
the scope of the present work.

\vspace*{1cm}

\section{Discussion}\label{secdiscussion}

In this paper, we have studied the thermodynamics of small-world
hypergraphs consisting  of sparse Poisson random graphs with
$p$-spin interactions superimposed onto a one-dimensional Ising
chain with nearest-neighbor interactions. Using a replica-symmetric
transfer-matrix analysis and the population dynamics algorithm we
have obtained the phase behavior of this system as a function of the
short-range and long-range couplings. We find for $p \geq 3$ that
all paramagnetic-ferromagnetic phase transitions are purely first
order, in contrast with $p=2$ where also a second order phase
transition occurs. For fixed $p$ and increasing connectivity $c$ the
ferromagnetic phase increases substantially and the transition line
converges to the analytically derived result for the $1+ \infty$
dimensional model. For the latter the ferromagnetic region decreases
for growing $p$. Using a bifurcation analysis we see that again,
$p=2$ has also a second-order transition and the first-order
transition occurs only for  $\beta J_0 \leq -0.275$.

\section*{Acknowledgments}
We would like to thank Heinz Horner for interesting discussions. We are indebted to the referee for pointing out an error in the first version of this manuscript. This work is partially supported by the Fund for Scientific Research Flanders-Belgium.

\begin{widetext}
\section*{Appendix: self-consistent equation for $\phi(x)$}

In order to derive the self-consistent equation for $\phi(x)$ we insert (\ref{uRS}) into the l.h.s. of (\ref{u}) using (\ref{T})
\begin{eqnarray}
\lefteqn{\sum_{\btau} T_{\bsigma,\btau}[F] u_0(\btau)} \label{selfconsistent1}
\nonumber \\
&=& \sum_{\btau} \exp \left( \beta J_0 \sum_{\alpha} \sigma^{\alpha} \tau^{\alpha} + c \sum_{\btau_1 \ldots \btau_{p-1}} \prod_{k=1}^{p-1} F(\btau_k) \left( e^{  \frac{\beta J}{c}\sum_{\alpha} \tau_1^{\alpha} \ldots \tau_{p-1}^{\alpha} \sigma^{\alpha} } -1 \right)\right) \int dx\: \phi(x) e^{\beta x \sum_{\alpha=1}^n \tau^\alpha}
\label{aa1}\\
&=& \sum_{\btau} \int dx\: \phi(x) e^{\beta x \sum_{\alpha=1}^n \tau^\alpha} e^{\beta J_0 \sum_{\alpha} \sigma^{\alpha} \tau^{\alpha}} e^{-c} \sum_{\mu=0}^{\infty} \left(\frac{c^{\mu}}{\mu!} \left(\sum_{\btau_1 \ldots \btau_{p-1}} \prod_{k=1}^{p-1} F(\btau_k) e^{  \frac{\beta J}{c}\sum_{\alpha} \tau_1^{\alpha} \ldots \tau_{p-1}^{\alpha} \sigma^{\alpha} } \right)^{\mu} \right)
\label{aa2}\\
&=& \left(\prod_{\alpha=1}^n \sum_{\gamma=\pm 1} \int dx\: \phi(x) e^{\beta x \gamma + \beta J_0  \sigma^{\alpha} \gamma}\right) \sum_{\mu=0}^{\infty} \left(\frac{e^{-c}c^{\mu}}{\mu!} \sum_{\{\btau_k^{\nu}\}_{k\leq p-1}^{\nu \leq \mu}} \prod_{\nu=1}^{\mu} \left[\prod_{k=1}^{p-1} F(\btau_k^{\nu})\right] e^{  \frac{\beta J}{c}\sum_{\alpha} \tau_1^{\nu,\alpha} \ldots \tau_{p-1}^{\nu,\alpha} \sigma^{\alpha} } \right)
\label{aa3}
\end{eqnarray}
In the transition to (\ref{aa2}) we have used that $F(\bpsi)$ is normalised to separate the term $e^{-c}$. We then have expanded the outermost of the remaining double exponential into a series. In (\ref{aa3}) we have written the powers as a product over a new replica index $\nu$. The vectors $\btau$ now have two replica indices. We also note that a poissonian factor appears. At this point we insert Eq.~(\ref{FRS}) to obtain
\begin{eqnarray}
\sum_{\btau} T_{\bsigma,\btau}[F] u_0(\btau) &=& \left(\prod_{\alpha=1}^n \sum_{\gamma=\pm 1} \int dx\: \phi(x) e^{\beta x \gamma + \beta J_0  \sigma^{\alpha} \gamma}\right) \nonumber \\
&&\times \sum_{\mu=0}^{\infty} \left(\frac{e^{-c}c^{\mu}}{\mu!} \sum_{\{\btau_k^{\nu}\}_{k\leq p-1}^{\nu \leq \mu}} \prod_{\nu=1}^{\mu} \left[\prod_{k=1}^{p-1} \int dh_k^{\nu}\: W(h_k^{\nu}) \prod_{\alpha=1}^n \frac{e^{\beta h_k^{\nu} \tau_k^{\nu,\alpha}}}{2 \cosh (\beta h_k^{\nu})}\right] e^{  \frac{\beta J}{c}\sum_{\alpha} \tau_1^{\nu,\alpha} \ldots \tau_{p-1}^{\nu,\alpha} \sigma^{\alpha} } \right) \nonumber \\
&=& \sum_{\mu=0}^{\infty} \left(\frac{e^{-c}c^{\mu}}{\mu!} \left[ \prod_{\nu=1}^{\mu}\prod_{k=1}^{p-1} \int dh_k^{\nu}\: \frac{W(h_k^{\nu})}{(2 \cosh (\beta h_k^{\nu}))^n} \right] \int dx\: \phi(x)  \right. \nonumber \\
&&\times \left. \prod_{\alpha=1}^n \left( \sum_{\gamma=\pm 1}e^{\beta \gamma (x + J_0  \sigma^{\alpha})}\right)
\prod_{\nu=1}^{\mu} \sum_{\gamma_1 \ldots \gamma_{p-1}}
e^{\beta \sum_{k=1}^{p-1} h_k^{\nu} \gamma_k + \frac{\beta J}{c} \gamma_1 \ldots \gamma_{p-1} \sigma^{\alpha} } \right) \label{lastline1}
\end{eqnarray}
We now focus on the second line of the last equation
\begin{eqnarray}
\lefteqn{\prod_{\alpha=1}^n \left( \sum_{\gamma=\pm 1}e^{\beta \gamma (x + J_0  \sigma^{\alpha})}\right)
\prod_{\nu=1}^{\mu} \sum_{\gamma_1 \ldots \gamma_{p-1}}
e^{\beta \sum_{k=1}^{p-1} h_k^{\nu} \gamma_k + \frac{\beta J}{c} \gamma_1 \ldots \gamma_{p-1} \sigma^{\alpha}}} \nonumber \\
&=& \exp \left( \sum_{\alpha=1}^n \log \left( G^{R}_{\sigma^{\alpha}}(x,\{h_k^{\nu}\}) \right)\right) \nonumber\\
&=& \exp \left( \sum_{\alpha=1}^n \sum_{s=\pm 1} \frac{1}{2}(1+ s\sigma_{\alpha}) \log \left( G^{R}_{s}(x,\{h_k^{\nu}\}) \right)\right) \nonumber \\
&=& \exp \left( \frac{1}{2}\left(\sum_{s=\pm 1} s \log \left( G^{R}_{s}(x,\{h_k^{\nu}\}) \right) \right) \sum_{\alpha=1}^n \sigma_{\alpha} + \frac{n}{2} \sum_{s=\pm 1} \log \left( G^{R}_{s}(x,\{h_k^{\nu}\}) \right)\right) \label{exprewrite1}
\end{eqnarray}
with (recall equation (\ref{GRS}))
\begin{equation}
G^{R}_{\sigma}(x,\{h_k^{\nu}\}) =  \left( \sum_{\gamma=\pm 1}e^{\beta \gamma (x + J_0  \sigma)}\right)
\prod_{\nu=1}^{\mu} \sum_{\gamma_1 \ldots \gamma_{p-1}}
e^{\beta \sum_{k=1}^{p-1} h_k^{\nu} \gamma_k + \frac{\beta J}{c} \gamma_1 \ldots \gamma_{p-1} \sigma}
\end{equation}
We take the limit $n \rightarrow 0$ in (\ref{exprewrite1}) and replace the last line of (\ref{lastline1}) with this limit
\begin{eqnarray}
\sum_{\btau} T_{\bsigma,\btau}[F] u_0(\btau) &=& \sum_{\mu=0}^{\infty} \left(\frac{e^{-c}c^{\mu}}{\mu!} \left[ \prod_{\nu=1}^{\mu}\prod_{k=1}^{p-1} \int dh_k^{\nu}\: \frac{W(h_k^{\nu})}{(2 \cosh (\beta h_k^{\nu}))^n} \right] \int dx\: \phi(x)  \right. \label{Tcontinued1} \nonumber \\
&&\times \left.
\exp \left( \frac{1}{2}\left(\sum_{s=\pm 1} s \log \left( G^{R}_{s}(x,\{h_k^{\nu}\}) \right) \right) \sum_{\alpha=1}^n \sigma_{\alpha}\right) \right) \nonumber \\
&=& \int dx' \sum_{\mu=0}^{\infty} \left(\frac{e^{-c}c^{\mu}}{\mu!} \left[ \prod_{\nu=1}^{\mu}\prod_{k=1}^{p-1} \int dh_k^{\nu}\: \frac{W(h_k^{\nu})}{(2 \cosh (\beta h_k^{\nu}))^n} \right] \int dx\: \phi(x)  \right. \nonumber \\
&&\times \left.
\delta\left[ x' - \frac{1}{2 \beta}\left(\sum_{s=\pm 1} s \log \left( G^{R}_{s}(x,\{h_k^{\nu}\}) \right) \right) \right]
\exp \left( \beta x' \sum_{\alpha=1}^n \sigma_{\alpha}\right) \right)
\end{eqnarray}
This expression is now of the form (\ref{uRS}) and identifying terms leads to
\begin{eqnarray*}
\lambda_0 \phi(x') &=& \sum_{\mu=0}^{\infty} \left(\frac{e^{-c}c^{\mu}}{\mu!} \left[ \prod_{\nu=1}^{\mu}\prod_{k=1}^{p-1} \int dh_k^{\nu}\: W(h_k^{\nu}) \right]
\int dx\: \phi(x)
\delta\left[ x' - \frac{1}{2 \beta}\left(\sum_{s=\pm 1} s \log \left( G^{R}_{s}(x,\{h_k^{\nu}\}) \right) \right) \right] \right)
\end{eqnarray*}
where we have used additionally that $(2 \cosh (\beta h_k^{\nu}))^n \rightarrow 1$ when $n \rightarrow 0$.

This equation can be simplified further as follows
\begin{eqnarray}
G^{R}_{s}(x,\{h_k^{\nu}\}) &=& \left( \sum_{\gamma=\pm 1}e^{\beta \gamma (x + J_0  \sigma)}\right)
\prod_{\nu=1}^{\mu} \sum_{\gamma_1 \ldots \gamma_{p-1}}
e^{\beta \sum_{k=1}^{p-1} h_k^{\nu} \gamma_k + \frac{\beta J}{c} \gamma_1 \ldots \gamma_{p-1} \sigma} \nonumber \\
&=& 4 \cosh \left(\beta (x + J_0  \sigma)\right)
\prod_{\nu=1}^{\mu} \sum_{\gamma_2 \ldots \gamma_{p-1}}
\cosh \left(\beta h_1^{\nu} + \frac{\beta J}{c} \gamma_2 \ldots \gamma_{p-1} \sigma \right) e^{\beta \sum_{k=2}^{p-1} h_k^{\nu} \gamma_k} \nonumber \\
&=& 4 \exp\left(\frac{1}{2} \log(\cosh(\beta h_1^\nu + \frac{\beta J}{c}) \cosh(\beta h_1^\nu - \frac{\beta J}{c}))\right) \cosh \left(\beta (x + J_0 \sigma )\right) \nonumber \\
&&\times \prod_{\nu=1}^{\mu} \sum_{\gamma_2 \ldots \gamma_{p-1}}
\exp \left(\gamma_2 \ldots \gamma_{p-1} \sigma\: \atanh (\tanh(\beta h_1^{\nu}) \tanh (\frac{\beta J}{c}))\right)
e^{\beta \sum_{k=2}^{p-1} h_k^{\nu} \gamma_k} \nonumber \\
&=& C(h_1^\nu, h_2^\nu, \ldots, h_{p-1}^\nu, J) \exp \left( \sum_{\nu=1}^{\mu} \sigma\: \atanh \left(\tanh (\frac{\beta J}{c}) \prod_{k=1}^{p-1} \tanh (\beta h_k^\nu)\right) \right)
\end{eqnarray}
with $C(h_1^\nu, h_2^\nu, \ldots, h_{p-1}^\nu, J)$ a function depending only on $h_1^\nu, h_2^\nu, \ldots, h_{p-1}^\nu, J$ and $\beta$. We finally arrive at
\begin{eqnarray}
\lefteqn{\frac{1}{2 \beta}\left(\sum_{s=\pm 1} s \log \left( G^{R}_{s}(x,\{h_k^{\nu}\}) \right) \right)} \nonumber \\
 &=& \frac{1}{\beta} \left[
\sum_{\nu=1}^{\mu} \atanh \left(\tanh (\frac{\beta J}{c}) \prod_{k=1}^{p-1} \tanh (\beta h_k^\nu)\right) + \atanh \left(  \tanh(\beta x) \tanh(\beta J_0) \right) \right] \label{phireduced}
\end{eqnarray}

In this way we have obtained the self-consistent equation (\ref{phi}) for $\phi(x)$. The equation for $\chi(x)$ can be derived in an analogous way.

From these equations we can also find the largest eigenvalue $\lambda_0$. Keeping the factor of order $n$ in (\ref{exprewrite1}) we end up with the following equation
\begin{eqnarray}
\lambda_0 \phi(x') &=& \sum_{\mu=0}^{\infty} \frac{e^{-c}c^{\mu}}{\mu!} \left[ \prod_{\nu=1}^{\mu}\prod_{k=1}^{p-1} \int dh_k^{\nu}\: \frac{W(h_k^{\nu})}{(2 \cosh (\beta h_k^{\nu}))^n} \right] \int dx\: \phi(x) \nonumber \\
&& \times \delta\left[ x' - \frac{1}{2 \beta}\left(\sum_{s=\pm 1} s \log \left( G^{R}_{s}(x,\{h_k^{\nu}\}) \right) \right) \right] \exp\left( \sum_{s = \pm 1} \frac{n}{2}\left(\log \left( G^{R}_{s}(x,\{h_k^{\nu}\}) \right)
\right) \right)
\end{eqnarray}
When we integrate both sides over $x'$ and expand the last exponential we find equation (\ref{lambda0}).

\end{widetext}


\begin{thebibliography}{99}

\bibitem{pekalski}
A. Pekalski, Phys. Rev. E {\bf 64}, 057104 (2001).

\bibitem{girvan}
M. Girvan and M.E.J. Newman, Proc. Natl. Acad. Sci. U.S.A. {\bf 99}, 7821 (2002).

\bibitem{simingli}
L. Siming \emph{et al}, Science {\bf 303}, 540 (2004).

\bibitem{barabasi_nature}
 A.-L. Barab\'asi and Z.N. Oltvai, Nat. Rev. Gen. {\bf 5}, 101 (2004).

\bibitem{newman}
M.E.J. Newman, Proc. Natl. Acad. Sci. U.S.A. {\bf 98}, 404 (2002).

\bibitem{barratw}
A. Barrat and M. Weigt,
Eur. Phys. J. B {\bf 13}, 547 (2000).

\bibitem{reptrans}
T. Nikoletopoulos, A. C. C. Coolen, I. P\'erez Castillo, N. S. Skantzos, J. P. L. Hatchett and B. Wemmenhove,
J. Phys. A: Math. Gen. {\bf  37}, 6455 (2004).


\bibitem{albert-barabasi}
R. Albert and A-L Barab\'asi, Rev. Mod. Phys. {\bf 74}, 47 (2002).

\bibitem{newman_review}
M.E.J. Newman, SIAM Rev. {\bf 45}, 167 (2003).

\bibitem{bookwatts}
Duncan J Watts,  \emph{Small Worlds: The Dynamics of Networks between Order and Randomness}  (Princeton University Press, Princeton, 2003).

\bibitem{bookdorogovtsev}
S.N. Dorogovtsev and J.F.F. Mendes, \emph{Evolution of Networks: From Biological Nets to the Internet and WWW} (Oxford University Press, London, 2003).

\bibitem{bookbarabasi}
A-L Barab\'asi, \emph{Linked: The New Science of Networks} (Oxford University Press, London, 2002).

\bibitem{ramadan}
E Ramadan, A Tarafdar, and A Pothen, \emph{18th International
Parallel and Distributed Processing Symposium (IPDPS'04) - Workshop
9    p. 189b}

\bibitem{wagner} A. Wagner and D.~A.~Fell, Proc. R. Soc. Lond. B
\textbf{268} 1803-1810 (2001)

\bibitem{ferroglass}
S. Franz, M. M\'ezard, F. Ricci-Tersenghi, M. Weigt, and R. Zecchina,
Europhys. Lett. {\bf 55}, 465 (2001).

\bibitem{barratz}
A. Barrat and R. Zecchina, Phys. Rev. E {\bf 59}, 1299 (1999).

\bibitem{ricci}
F. Ricci-Tersenghi, M. Weigt and R. Zecchina, Phys. Rev. E {\bf 63}, 026702 (2001).

\bibitem{heylen}
R. Heylen, N.S. Skantzos, J. Busquets Blanco and D. Boll\'e, Phys. Rev. E {\bf 73}, 016138 (2006).

\bibitem{mezardparisi}
M. M\'ezard and G. Parisi,
Eur. Phys. J. B {\bf 20}, 217 (2001).

\bibitem{1infty}
N.S. Skantzos and A.C.C. Coolen,
J. Phys. A: Math. Gen. {\bf  33}, 5785 (2000).

\bibitem{kardar}
M. Kardar,
Phys. Rev. {\bf 28}, 244 (1983).

\bibitem{MPV}
M. M\'ezard, G. Parisi and M.A. Virasoro, \emph{Spin Glass Theory and Beyond} (World Scientific, Singapore, 1987).

\end{thebibliography}
\end{document}